\documentclass[aps, prb, reprint, amsmath, amssymb, groupedaddress, footinbib]{revtex4-2}
\usepackage[colorlinks=true, allcolors=blue]{hyperref}
\usepackage{graphicx}

\newcommand{\lt}{\left}
\newcommand{\rt}{\right}

\newcommand{\bk}{\mathbf{k}}

\begin{document}

\title{Slow convergence of spin-wave expansion and magnon dispersion in the 1/3 plateau of the triangular XXZ antiferromagnet}

\author{Achille Mauri}
\author{Fr\'{e}d\'{e}ric Mila}%
\affiliation{%
Institute of Physics, Ecole Polytechnique F\'{e}d\'{e}rale de Lausanne (EPFL), CH-1015
Lausanne, Switzerland
}%

\date{\today}

\begin{abstract}
Motivated by recent experiments on the quantum magnet K$_2$Co(SeO$_3$)$_2$, we study theoretically
the excitation spectrum of the nearest-neighbour triangular XXZ model in the limit of strong
easy-axis anisotropy, within the “up-up-down” (1/3-plateau) phase. We make an expansion in $\alpha
=J_{xy}/J_{zz}$ instead of $1/S$ and calculate the magnon dispersion for any value of the spin $S$
at second order in $\alpha$, with two important conclusions: (i) the $1/S$ expansion converges very
slowly for $S=1/2$, making spin-wave theory quantitatively inaccurate up to very large orders; (ii)
compared to the linear spin-wave predictions, our magnon dispersion presents a much better agreement
with experimental results on K$_2$Co(SeO$_3$)$_2$, for which $\alpha \simeq 0.08$.
\end{abstract}

\maketitle

The nearest-neighbour spin-$1/2$ XXZ antiferromagnet on the triangular lattice is one of the
prototype models in the theory of geometrically-frustrated quantum magnetism, and has been a
subject of investigation for decades.
In the case in which the exchange anisotropy is of easy-axis type, an early
analysis~\cite{fazekas_pm_1974} suggested the possibility that the model may host a quantum spin
liquid ground state, but later studies did not corroborate this scenario and predicted instead an
ordered state at zero field, followed by a sequence of transitions between ordered phases in
presence of an external magnetic field directed along the $c$ axis (the easy axis of the exchange
anisotropy)~\cite{kleine_zpb_1992, kleine_zpb_1992b, sen_prl_2008, wang_prl_2009, jiang_prb_2009,
heidarian_prl_2010, zhang_prb_2011, yamamoto_prl_2014, sellmann_prb_2015, yamamoto_prb_2019,
xu_arxiv_2024}.
For zero magnetic field, in particular, different theoretical analyses predicted a
``spin-supersolid'' phase~\cite{sen_prl_2008, wang_prl_2009, jiang_prb_2009,heidarian_prl_2010,
zhang_prb_2011,  yamamoto_prl_2014, sellmann_prb_2015, yamamoto_prb_2019,
xu_arxiv_2024}, with simultaneous breaking of a $Z_{3}$ translational invariance and the $U(1)$
rotational symmetry.
Some results reported recently, however, gave indications that the ground state may be a ``solid''
with unbroken $U(1)$ symmetry~\cite{ulaga_arxiv_2024, ulaga_prb_2024} (see also
Ref.~\cite{xu_arxiv_2024}).
In presence of a $c$-axis oriented field, the zero-temperature phase diagram has been
predicted to present, in addition to the low-field phase, a $1/3$-plateau phase, a high-field
supersolid, and a fully saturated phase~\cite{yamamoto_prl_2014, yamamoto_prb_2019, gao_npj_2022,
xu_arxiv_2024}.

The quantum dynamics of the easy-axis $S = 1/2$ XXZ model has recently received interest in
connection with experiments on the compounds Na$_{2}$BaCo(PO$_{4}$)$_{2}$~\cite{zhong_pnas_2019,
gao_npj_2022, jia_prr_2024, chi_prb_2024, sheng_arxiv_2024}, NdTa$_{7}$O$_{19}$~\cite{arh_nm_2022,
ulaga_prb_2024, ulaga_arxiv_2024}, and K$_{2}$Co(SeO$_{3}$)$_{2}$~\cite{zhong_prm_2020,
zhu_prl_2024, chen_arxiv_2024} (in short ``KCSO").
In the case of KCSO, in particular, experimental studies were reported recently by two independent
groups~\cite{zhu_prl_2024, chen_arxiv_2024}, who analyzed the phase diagram and the excitations in
this compound by a combination of thermodynamic, magnetic, and neutron scattering experiments.
The experimental results showed that the easy-axis anisotropy in this material is very strong,
implying that the effective model describing the compound is close to the Ising limit.
The magnetization curve and the field-temperature phase diagram observed experimentally were found
to be consistent with a theoretical interpretation in terms of  two supersolid phases and a $1/3$
plateau phase.
At the same time, inelastic neutron scattering (INS) showed that the excitation spectrum presents
features which contrast sharply with the single-magnon spectrum predicted by linear spin wave theory
(LSWT).
At zero magnetic field, the contrast is particularly striking: INS reveals a deep minimum near the M
point, which is completely absent in LSWT, and broad spectral features, which conflict with
an interpretation in terms of single-magnon modes.
The origin of these observations is currently under debate~\cite{zhu_prl_2024, chen_arxiv_2024,
xu_arxiv_2024, ulaga_arxiv_2024}.

Within the $1/3$-plateau phase, the excitation spectra measured by INS at magnetic fields $H = 1.5$
T~\cite{zhu_prl_2024} and $B = 7$ T~\cite{chen_arxiv_2024} revealed instead a negligible broadening
(smaller than the experimental resolution), implying that the spectral function is dominated
by sharp single-magnon excitations.
However, even in this case, the analysis of Ref.~\cite{zhu_prl_2024} showed that it is not
possible to obtain a fully satisfactory interpretation of the INS data within LSWT and assuming the
nearest-neighbour XXZ model.
In fact, it was shown that the magnon dispersions predicted by LSWT, when calculated assuming the
values of the exchange constants deduced from the magnetization vs. field curve, present a
discrepancy to the experimental dispersions, in particular in the curvature of one of the modes
near the $\Gamma$ and K points.

The authors of Ref.~\cite{zhu_prl_2024} suggested that the observed discrepancies arise from
corrections beyond LSWT.
Effects beyond the spin-wave approximation in fact have been investigated in the context of the
compound Ba$_{3}$CoSb$_{2}$O$_{9}$ and it has been shown theoretically that in the $uud$ phase of
this material the corrections to LSWT are strong, despite the collinearity of the magnetic
order~\cite{kamiya_nc_2018}.
We also note that corrections to LWST have been reported for the $uud$ phase of
Na$_{2}$BaCo(PO$_{4}$)$_{2}$~\cite{sheng_arxiv_2024}.

In this Letter, motivated by the recent experiments on KCSO, we analyze the magnon dispersion in
the $1/3$ plateau phase of the XXZ model using an approach beyond LSWT.
To address this question, we use a perturbative expansion up to second order in the
ratio $\alpha = J_{xy}/J_{zz}$ between the transverse exchange $J_{xy}$ and the longitudinal
(Ising-like) exchange $J_{zz}$, an expansion which we expect to be quantitatively 
accurate in KCSO, due to the strong easy-axis anisotropy in this material.

The starting point of our analysis is the $S  = 1/2$ triangular XXZ model, described by the
Hamiltonian:
\begin{equation} \label{H}
\begin{split}
{\cal H} & = \sum_{\langle i, j \rangle} \lt(J_{zz} S^{z}_{i} S^{z}_{j} + J_{xy} (S^{x}_{i}
S^{x}_{j} + S^{y}_{i} S^{y}_{j})\rt) \\
& \quad - g_{z} \mu_{\rm B} H \sum_{i} S^{z}_{i}~,
\end{split}
\end{equation}
where the sum $\sum_{\langle i, j \rangle}$ runs over the nearest-neighbour bonds of
the triangular lattice, and the term $- g_{z} \mu_{\rm B}H \sum_{i} S^{z}_{i}$ encodes the coupling
to a longitudinal ($c$-axis oriented) field $H$.

\begin{figure}
\centering
\includegraphics[scale=1]{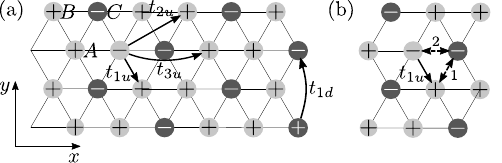}
\caption{\label{UUD_configuration} (a) $uud$ configuration and magnon hoppings in second-order
perturbation theory.
At order $\alpha^{2}$, the $u$ excitations are described by an effective tight binding problem with
hopping amplitudes $t_{1u}$, $t_{2u}$, and $t_{3u}$ on the honeycomb lattice.
The $d$ excitations, at the same order, are described by an effective hopping problem with amplitude
$t_{1d}$ between the nearest-neighbour C sites (which are second-nearest neighbours on the full
lattice).
(b) Perturbative process leading to the second-order correction to $t_{1u}$: two pairs of spins are
interchanged sequentially (flipping pair '1', then pair '2') resulting in the motion of the magnon
excitation. $t_{2u}$, $t_{3u}$, and $t_{1d}$ are determined by similar processes.
}
\end{figure}

In the case of KCSO, Zhu \emph{et al.}~deduced from the experimental magnetization vs. field curve
the values of the exchange constants and of the gyromagnetic ratio $J_{zz} = 3.1(1)$ meV, $\alpha =
J_{xy}/J_{zz} = 0.08(1)$, $g_{z} \simeq 7.9$~\cite{zhu_prl_2024}.
These values are consistent with those reported by Chen~\emph{et al.}~\cite{chen_arxiv_2024}.
In particular, we note that by fitting the INS spectrum at 7 T to linear spin wave theory
Ref.~\cite{chen_arxiv_2024} extracted the exchange constants $J_{zz} = 2.98(2)$ meV, $J_{xy} =
0.21(3)$  meV [corresponding to an anisotropy ratio $\alpha = 0.070(15)$].
The small values of $\alpha$ thus justify quantitatively a perturbation expansion in $\alpha$.

Strictly in the Ising limit $J_{xy} = 0$ and for zero temperature, the model presents two
phases: an ``up-up-down'' ($uud$) phase in the interval of fields  $0 < g_{z} \mu_{\rm B} 
H < 3 J_{zz}$, and a fully saturated phase for $g_{z} \mu_{\rm B} H > 3
J_{zz}$~\cite{zhang_prb_2011}.
In this Letter, we focus entirely on the $uud$ phase, which corresponds to the $1/3$ magnetization
plateau observed experimentally in KCSO~\cite{zhu_prl_2024, chen_arxiv_2024}.
The structure of the $uud$ ground state is shown in Fig.~\ref{UUD_configuration}.
A triangular sublattice (sublattice C) is occupied by down spins (with $S^{z} = -1/2$) 
while the two remaining sublattices (A and B) which, together, form a honeycomb lattice are
occupied by up spins.

The magnon excitations on top of the $uud$ state are gapped and correspond to single spin flips.
In the Ising limit, a spin flip at one of the sites of the C sublattice (called in the following
"$d$" excitation) has an excitation gap $\epsilon^{(0)}_{d} = 3 J_{zz} - g_{z} \mu_{\rm B}
H$, while spin flips at the A and B sites ("$u$" excitations), have excitation gap
$\epsilon_{u}^{(0)} = g_{z} \mu_{\rm B}H$.

Turning on a small $J_{xy}$ leads to a perturbative dressing of the $uud$ ground state
and induces a dispersion of the magnon excitations~\cite{chen_arxiv_2024}.
For the $u$ excitations, the dispersion at second order in $\alpha$ is described by an effective
hopping problem on the honeycomb lattice, with three non-zero hopping amplitudes between first,
second, and third nearest neighbours (see Fig.~\ref{UUD_configuration}).
By an explicit calculation, we find that the correspond
\footnotetext[2]{A. Mauri and F. Mila, 'Spin-wave expansion and magnon dispersion in the 1/3
plateau of the triangular XXZ antiferromagnet'. Zenodo, 2025.
doi: \href{https://doi.org/10.5281/zenodo.15270189}{https://doi.org/10.5281/zenodo.15270189}}ing
hopping amplitudes are $t_{1u} =
-\alpha(1 - \alpha/2)J_{zz}/2$, $t_{2u}=t_{3u} = \alpha^{2}J_{zz}/4$.
In addition, perturbative corrections give a momentum-independent energy shift, which arises from
the disturbance induced by an excitation on the virtual transitions occurring in 
its proximity.
This leads to a renormalization of the gap, which can be encoded in an energy shift $V_{u} = -
5\alpha^{2} J_{zz}/8$.
We note that the effective parameters $t_{1u}$, $t_{2u}$, $t_{3u}$, $V_{u}$ which we derived are
equal, in a different notation, to those obtained in Ref.~\cite{zhang_prb_2011} for an
equivalent problem of hard-core bosons.

In momentum space, the solution of the single-particle hopping problem allows us to calculate the
exact magnon dispersions at order $\alpha^{2}$, which read (see App.~A):
\begin{equation} \label{epsilon_2_u}
\begin{split}
\epsilon^{(2)}_{u, \pm}(\bk) & = g_{z} \mu_{\rm B} H - \frac{\alpha^{2}}{8} J_{zz} (2 
|f_{\bk}|^{2} -
1) \\
& \qquad \pm \frac{\alpha J_{zz}}{2} \lt|\lt(1 + \frac{\alpha}{2}\rt) f_{\bk} - \frac{\alpha}{2}
f^{*2}_{\bk}\rt|~,
\end{split}
\end{equation}
with $f_{\bk} = {\rm e}^{i k_{x}} + {\rm e}^{-\frac{i}{2}(k_{x} + \sqrt{3}k_{y})} + {\rm e}^{-
\frac{i}{2} (k_{x} - \sqrt{3} k_{y})}$.

For the $d$ excitations, an analogue analysis at second order shows that the magnon 
spectrum is described by an effective hopping problem on the C sublattice with a
nearest-neighbour hopping amplitude $t_{1d} = \alpha^{2} J_{zz}/2$ and an energy shift
$V_{d} = - 3 \alpha^{2} J_{zz}/4$.
The resulting dispersion relation, exact at order $\alpha^{2}$, reads:
\begin{equation} \label{epsilon_2_d}
\epsilon^{(2)}_{d}(\bk) = 3 J_{zz} - g_{z} \mu_{\rm B} H - \frac{\alpha^{2}}{4} J_{zz} (2 
|f_{\bk}|^{2}
- 3)~.
\end{equation}

In contrast with Eq.~\eqref{epsilon_2_u},~\eqref{epsilon_2_d}, which are exact for $S = 1/2$, the
spectrum computed within the linear spin-wave approximation is given by $\epsilon_{a}^{({\rm
LSWT})}(\bk) = |\lambda_{a}(\bk)|$, where $\lambda_{i}(\bk)$ are the eigenvalues of:
\begin{equation}
{\cal M}^{({\rm LSWT})}(\bk) = g_{z} \mu_{\rm B}H I + S J_{zz}
\begin{vmatrix}
0 & \alpha f_{\bk} & \alpha f^{*}_{\bk} \\
\alpha f^{*}_{\bk} & 0 & \alpha f_{\bk} \\
- \alpha f_{\bk} & - \alpha f^{*}_{\bk} & - 6
\end{vmatrix}~,
\end{equation}
and $I$ is the $3 \times 3$ identity matrix.
Expanding the eigenvalues for small $\alpha$, we find that the spectrum of LSWT, at order
$\alpha^{2}$ is~(see App.~B):
\begin{equation} \label{epsilon_LSWT}
\begin{split}
\epsilon^{(2, {\rm LSWT})}_{u, \pm} (\bk) & = g_{z} \mu_{\rm B} H - \frac{S}{6} 
\alpha^{2} J_{zz}
|f_{\bk}|^{2} \\
& \qquad \pm S \alpha J_{zz} \lt|f_{\bk} - \frac{\alpha}{6} f^{*2}_{\bk}\rt|~, \\
\epsilon^{(2, {\rm LSWT})}_{d}(\bk) & =  6 S J_{zz} - g_{z} \mu_{\rm B} H -
\frac{S}{3}
\alpha^{2} J_{zz} |f_{\bk}|^{2}~.
\end{split}
\end{equation}

Comparing these expressions with Eqs.~\eqref{epsilon_2_u} and~\eqref{epsilon_2_d} we can see that,
for $S = 1/2$, the LSWT approximation reproduces correctly only the leading terms of the exact
dispersions (up to order $\alpha$), but misses severely the coefficients of the next-to-leading
corrections, of order $\alpha^{2}$, in comparison to the exact $S = 1/2$ results.
Although $\alpha$ is small, the next-to-leading corrections appear multiplied with 
factors of order $|f_{\bk}|$, which become $|f_{\bk}| \simeq 3$ near the $\Gamma$ and the 
K point.
As a result, even for the small value of the anisotropy ratio characteristic of KCSO, $\alpha =
0.08(1)$, the corrections are sizeable.

\begin{figure}
\centering
\includegraphics{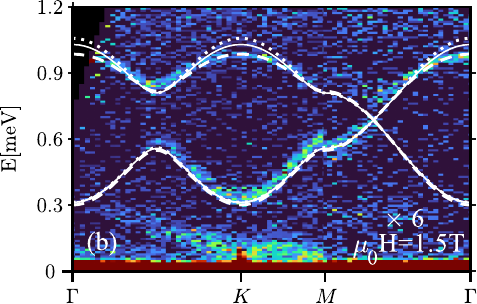}
\caption{\label{spectrum_UUD_plateau} Magnon dispersions $\epsilon_{u, \pm}(\bk)$ in first- and
second-order perturbation theory (dotted and dashed lines, respectively), and LSWT dispersions
(solid lines) plotted in superposition to Fig.~4b of Ref.~\cite{zhu_prl_2024} (adapted with
permission from the authors).
All dispersions are calculated assuming as parameters $J_{zz} = 3.1 $ meV, $\alpha = 0.08$, and
$g_{z} = 7.9$.
The solid lines, illustrating the linear-spin wave dispersions, fall on top of the LSWT
spectra shown in Fig.~4b of Ref.~\cite{zhu_prl_2024}.
}
\end{figure}

For example, one of the prominent effects of the next-to-leading corrections (due to the 
hopping $t_{2u}$) is that they introduce an asymmetry between the bandwidths of the two 
branches $\epsilon_{u, -}(\bk)$ and $\epsilon_{u, +}(\bk)$, lifting the reflection 
symmetry of the first-order spectrum $\epsilon^{(1)}_{u,\pm}(\bk) = g_{z} \mu_{\rm B} H 
\pm \alpha J_{zz} |f_{\bk}|/2$.
Eq.~\eqref{epsilon_2_u} implies that the ratio between the bandwidths $W_{+}$ and $W_{-}$ of the
branches $\epsilon_{\pm}$, including the next-to-leading corrections is $W_{+}/W_{-} = (1 - 5
\alpha/2)/(1 + \alpha/2) + O(\alpha^{2}) = 1 - 3 \alpha + O(\alpha^{2})$.
The induced asymmetry is strongly underestimated by LSWT, which predicts a spectrum with $W_{+}^{\rm
(LSWT)}/W_{-}^{\rm (LSWT)} = 1 - \alpha + O(\alpha^{2})$ (missing the asymmetry by a factor $3$ in
the limit of small $\alpha$).
The hopping $t_{3u}$ (also underestimated by a factor $3$ by LSWT for $\alpha \to 0$) has 
the more subtle effect of modifying the dispersions without affecting the reflection 
symmetry of $\epsilon_{u \pm}(\bk)$.

Due to these large corrections to LSWT, the exact second-order formula~\eqref{epsilon_2_u} provides
a useful tool for revisiting the interpretation of the experimental INS results.
To compare the theoretical predictions with the experimental spectrum, in
Fig.~\ref{spectrum_UUD_plateau} we show the dispersions predicted by LSWT (solid lines),
first-order perturbation theory (dotted lines) and second-order perturbation theory (dashed lines),
plotted in superposition onto the INS data of Ref.~\cite{zhu_prl_2024}.
The experimental spectrum, which was measured in a field $H = 1.5$ T, shows clearly two 
low-energy magnon modes, which in our notation correspond to $\epsilon_{u, \pm}(\bk)$.
The third branch $\epsilon_{d}(\bk)$, instead, was not observed, which can be explained from the
fact that its energy lied outside the experimental energy range (the third branch, however, was
observed at $B = 7$ T in Ref.~\cite{chen_arxiv_2024}).
In Fig.~\ref{spectrum_UUD_plateau} all theoretical lines are calculated using 
the parameters $J_{zz} = 3.1$ meV, $\alpha = 0.08$, $g_{z} = 7.9$, which 
Ref.~\cite{zhu_prl_2024} deduced from the magnetization curve.
The LSWT prediction for $\epsilon_{u, \pm}$, as it was noted in Ref.~\cite{zhu_prl_2024}, 
presents a discrepancy to the data, visible particularly in that it overestimates the 
curvature of $\epsilon_{u, +}(\bk)$ near the $\Gamma$
and the K points.
The second-order perturbative formula, Eq.~\eqref{epsilon_2_u}, instead, is seen 
to present a much closer agreement.

We note that Zhu~\emph{et al.} presented an alternative analysis of the data by fitting 
LSWT to the spectrum~\cite{zhu_prl_2024, Note1}.
By this analysis it was shown that the LSWT dispersions $\epsilon_{u\pm}(\bk)$ could be matched
well to the experimental results.
However, this required a set of fitting parameters [$\tilde{J}_{zz} = 0.68(6)$ meV, $\tilde{J}_{xy}
= 0.25(1)$ meV, $\tilde{g} = 8.34(6)$] in which the value of $\tilde{J}_{zz}$ is more 
than $4$ times smaller than the bare $J_{zz}$.
This values of $\tilde{J}_{zz}$ are in sharp contrast with other complementary 
experimental observations~\cite{Note1}.
The second-order formula, instead, provides a closer agreement when calculated in terms 
of the bare exchange constants, without any adjustment of the parameters.

There remains some tension between the INS and the predictions of Eq.~\eqref{epsilon_2_u}.
Considering for example the bandwidth asymmetry we can estimate from the fit of 
Ref.~\cite{zhu_prl_2024} that the experimental ratio between the bandwidths of the 
$\epsilon_{u, \pm}$ branches is $W_{+}/W_{-} = 0.67(4)$.
The ratio $W_{+}/W_{-} = (1- 5\alpha/2)/(1 + \alpha/2)$ implied by Eq.~\eqref{epsilon_2_u}, instead,
evaluates to $W_{+}/W_{-} = 0.77(3)$ assuming $\alpha = 0.08(1)$.
This residual discrepancy may be due to higher-order terms in the $\alpha$ expansion.

In addition, the residual discrepancy could in part arise from longer-range interactions 
beyond the nearest-neighbour exchanges considered in the XXZ model.
An additional second-nearest-neighbour transverse exchange $J_{2xy}$ contributes to the 
asymmetry of the spectrum, giving an additional term to $W_{+}/W_{-}$ equal in 
first order to $\Delta(W_{+}/W_{-}) = 6 J_{2xy}/J_{xy}$.
Thus, we can estimate that if $J_{2xy}$ is really the origin of the deviation in 
$W_{+}/W_{-}$ then a very small ferromagnetic $J_{2xy} \approx - 0.015(10) J_{xy}$ would 
be enough to compensate for the difference between the theoretical and the 
experimental $W_{+}/W_{-}$.
We note, however, that if the same analysis had been done within LSWT, the value of the 
effective $t_{2u} = \alpha^{2} J_{zz}/4$ would have been replaced with its LSWT 
approximation $t_{2u} = \alpha^{2}J_{zz}/12$ and, the difference between the 
two hoppings, $\alpha^{2} J_{zz}/6$, would have been incorrectly ascribed to a physical 
exchange constant $J_{2xy}$, leading to an error $\Delta J_{2xy} = -\alpha^{2} J_{zz}/3 
\approx -0.026(3) J_{xy}$.
Similar considerations apply to third-nearest-neighbour interactions and to Ising 
couplings like $J_{2zz}$.
This effect may give some corrections to the bounds on $J_{2xy}$, $J_{2zz}$ found in 
Ref.~\cite{chen_arxiv_2024}.

Finally, we note that the expansion near the Ising limit offers the possibility to assess not only
the quality of the LSWT approximation, but also the convergence of the semiclassical $1/S$ series.
This can be controlled by studying the generalization of the problem to an arbitrary
value of the spin $S$.
To repeat the analysis, the $uud$ ground state configuration is promoted to a state in which
two sublattices (A and B) are occupied by spins $+S$, whereas the remaining sublattice C is occupied
by spins $-S$, and the $u$ and $d$ excitations, become the lowering or raising by $1$ of 
the spins on the A, B or on the C sublattices.
At order $\alpha^{2}$, the effective hopping problem on the honeycomb lattice which 
governs the $u$ excitations has the same hopping amplitudes as in the $S = 1/2$ case,
but with a nontrivial dependence on the spin $S$:
\begin{equation}\label{hopping_matrix_elements}
\begin{gathered}
V_{u}(S)  = -\frac{S (9 S^{2} - 6 S + 2)}{(3S-1) (6S - 1)} \alpha^{2} 
J_{zz}~, \\
t_{1u}(S)  = - S \alpha \lt(1 - \frac{2S \alpha}{6 S - 1}\rt) J_{zz}~, \\
t_{2u}(S)  = t_{3u}(S) = \frac{S^{2}}{2(3 S - 1)} \alpha^{2} J_{zz}~. \\
\end{gathered}
\end{equation}

For the $d$ excitations, similarly, the energy shift $V_{d}$ and the hopping amplitude
$t_{1d}$ on the C sublattice are:
\begin{equation}
\begin{gathered}
V_{d}(S)  = - \frac{3 (6 S^{2} - 4 S + 1)S}{(6S-1)(3S-1)} 
\alpha^{2}
J_{zz} \\
t_{1d}(S)  = \frac{S^{2}}{3 S - 1} \alpha^{2} J_{zz}~.
\end{gathered}
\end{equation}

\begin{figure}
 \centering
 \includegraphics[scale=1]{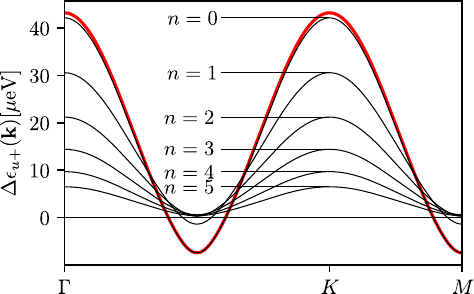}
 \caption{\label{convergence_large_S} Difference $\Delta \epsilon_{u+}(\bk) = \epsilon_{u+}(\bk) -
\epsilon^{(2)}_{u+}(\bk)$ between the magnon dispersion $\epsilon_{u+}(\bk)$ and the exact $S =
1/2$ second-order dispersion $\epsilon^{(2)}_{u+}(\bk)$ calculated within LSWT (red solid line) and
in the $1/S$ expansion (black solid lines), assuming as parameters $J_{zz}= 3.1$ meV, $\alpha =
0.08$.
The black solid lines show the ``error'' $\Delta \epsilon_{u+}(\bk)$ committed by truncating at
order $1/S^{n-1}$ the large-$S$ expansion of Eq.~\eqref{epsilon_2_u_any_S} (so that $n = 0$
corresponds to the large-$S$ limit).
The difference between the red curve and the $n = 0$ line is due to corrections beyond order
$\alpha^{2}$, which are included in LSWT, but not in the $S \to \infty$ limit of
Eq.~\eqref{epsilon_2_u_any_S}.}
\end{figure}

The resulting dispersions,
\begin{equation}\label{epsilon_2_u_any_S}
\begin{split}
& \epsilon_{u, \pm}^{(2)}(\bk)  = g_{z} \mu_{\rm B} H - \frac{S \alpha^{2} J_{zz}}{2 (3S -
1)} \lt(S |f_{\bk}|^{2} - \frac{9S - 4}{6 S - 1}\rt) \\
& \pm S \alpha J_{zz} \lt|f_{\bk} + \frac{S \alpha f_{\bk}}{(3S - 1) (6 S - 1)} - \frac{S
\alpha f^{*2}_{\bk}}{2 (3S - 1)}\rt|~,
\end{split}
\end{equation}
\begin{equation}
\begin{split}
\epsilon_{d}^{(2)}(\bk) & = 6 S J_{zz} -  g_{z} \mu_{\rm B} H + \frac{3S}{6 S - 1} 
\alpha^{2} J_{zz} \\
& - \frac{S^{2}}{3S - 1} \alpha^{2} J_{zz} |f_{\bk}|^{2}~,
\end{split}
\end{equation}
interpolate between the exact spin-$1/2$ results and the linear spin wave limit ($S \to 
\infty$).
In addition to providing a result for general $S$ which may be useful on its own, 
these expressions allow to control the convergence of the nonlinear spin-wave ($1/S$) 
expansion for $S = 1/2$.
To analyze the convergence, Fig.~\ref{convergence_large_S} shows the difference between the
dispersion $\epsilon^{(2)}_{u+}(\bk)$ truncated at order $n$ in the large-$S$ expansion 
and Eq.~\eqref{epsilon_2_u} (exact for spin $S = 1/2$ at order $\alpha^{2}$).
Interestingly, this difference converges to zero very slowly with the order $n$.
This can be traced to the very slow convergence of denominators such as $1/(3S - 1) \simeq 1/(3S)
+1/(9 S^{2}) + ... $ that appear in the $\alpha$ correction.
As a result, it would have been essentially impossible to obtain an accurate approximation by
truncating the semiclassical expansion at a low order.

To conclude, we derived exact dispersion relations up to second order in $\alpha = 
J_{xy}/J_{zz}$ for the magnon excitations on top of the $1/3$ plateau phase of the XXZ 
model.
The analysis shows that the next-to-leading corrections of order $\alpha^{2}$ have a  
dramatically different magnitude in comparison with their approximate predictions in 
LSWT, and also with the predictions of a low-order $1/S$ expansion.
Whether a similarly slow convergence of the $1/S$ series persists in different models with 
more isotropic interactions remains an important open question.
For the magnetic excitations of KCSO, our analysis corroborates the validity of the XXZ 
model with only nearest-neighbour exchange, and provides a framework to constrain with 
higher accuracy the magnitude of possible (small) longer-range couplings.

\emph{Acknowledgments}---We acknowledge useful discussions with Andrey Zheludev.
We also thank Cristian Batista, Alexander Chernyshev, Andreas L\"{a}uchli, and Oleg Starykh for
useful discussions on related topics.
This work has been supported by the Swiss National Science Foundation Grant No. 212082.

\emph{Data availability}---The data that support the findings of this article are
available~\cite{Note2}.

\appendix

\section*{Appendix A: Magnon dispersion relations}

The spectrum of the ``$u$'' excitations is determined by a single-particle hopping problem on
the honeycomb lattice, whose Hamiltonian can be written as:
\begin{equation}
\begin{split}
{\cal H}_{u} & =  (g_{z} \mu_{\rm B} H + V_{u}) \sum_{i} a^{+}_{i} a_{i} \\
& - \sum_{i, j} (t_{1u} I_{1 ij} + t_{2u} I_{2 ij} + t_{3u} I_{3 ij}) a^{+}_{i} a_{j}~,
\end{split}
\end{equation}
where $I_{1ij}$, $I_{2ij}$, and $I_{3 ij}$ are matrices connecting, respectively, first, second,
and third nearest-neighbours.
(In this definition $I_{nij} = 1$ if $i$ and $j$ are $n$th nearest-neighbours on the honeycomb
lattice and zero otherwise).
In momentum space, the Hamiltonian reads:
\begin{equation}
{\cal H}_{u} = \sum_{\bk} \begin{pmatrix} a^{+}_{A\bk}&  a^{+}_{B\bk} \end{pmatrix}
h(\bk)
\begin{pmatrix}
 a_{A\bk} \\
 a_{B \bk}
\end{pmatrix}
\end{equation}
where $a_{{\rm A}\bk}^{+}$, $a_{{\rm B}\bk}^{+}$, $a_{{\rm A}\bk}$, $a_{{\rm B}\bk}$ are
creation/annihilation operators of plane wave states,
\begin{equation}
\begin{split}
h(\bk) & = \begin{vmatrix}
g_{z} \mu_{\rm B} H + V_{u} & -t_{1u} f_{\bk} \\
-t_{1u} f^{*}_{\bk} & g_{z}\mu_{\rm B} H + V_{u}
         \end{vmatrix}\\
& - \begin{vmatrix}
t_{2u} (|f_{\bk}|^{2} - 3) &  t_{3u} (f^{*2}_{\bk} -2f_{\bk}) \\
t_{3u} (f^{2}_{\bk} - 2 f^{*}_{\bk}) & t_{2u} (|f_{\bk}|^{2} - 3)
    \end{vmatrix}
\end{split}
\end{equation}
and, as in the main text, $f_{\bk} = {\rm e}^{i k_{x}} + {\rm e}^{-\frac{i}{2} (k_{x} + \sqrt{3}
k_{y})} + {\rm e}^{- \frac{i}{2} (k_{x} - \sqrt{3} k_{y})}$.

The spectrum at second-order is then
\begin{equation}
\begin{split}
\epsilon^{(2)}_{u\pm}(\bk) & = g_{z} \mu_{\rm B} H + V_{u} - t_{2u} (|f_{\bk}|^{2} - 3)\\
& \pm \lt|(t_{1u} - 2 t_{3u}) f_{\bk} + t_{3u} f^{*2}_{\bk} \rt|~.
\end{split}
\end{equation}
Replacing $V_{u}$, $t_{1u}$, $t_{2u}$, $t_{3u}$ with Eqs.~\eqref{hopping_matrix_elements} gives the
dispersions presented in the main text.

The ``$d$``-type excitations hop among the sites of the C sublattice, which is a triangular Bravais
lattice.
The corresponding hopping problem in real space reads ${\cal H}_{d} = \sum_{i} (6 S 
J_{zz} - g_{z} \mu_{\rm B} H + V_{d} ) a^{+}_{i}
a_{i} - t_{1d} \sum_{i, j} I_{2ij} a^{+}_{i} a_{j}$ and the dispersion relations,
\begin{equation}
\epsilon_{d}^{(2)}(\bk) = 6 S J_{zz} - g_{z} \mu_{\rm B} H + V_{d} - t_{1d} (|f_{\bk}|^{2} 
- 3)~.
\end{equation}

As a remark, we note that the explicit values of the hoppings 
[Eq.~\eqref{hopping_matrix_elements} in the main text] have the property that $t_{2u} = 
t_{3u}$ for any $S$.
This implies that the dispersions $\epsilon^{(2)}_{u, \pm}(\bk)$ can be written as
\begin{equation}
\begin{split}
\epsilon^{(2)}_{u, \pm}(\bk) & = g_{z} \mu_{\rm B} H + V_{u} + t_{1u} \Big[-\gamma 
(|f_{\bk}|^{2} -3) \\
& \pm \lt|(1 - 2 \gamma ) f_{\bk} + \gamma f^{*2}_{\bk}\rt|\Big]~,
\end{split}
\end{equation}
with $\gamma = t_{2u}/t_{1u} = t_{3u}/t_{1u} \simeq - S \alpha/(3S-1)$.
This relation shows that up to an energy shift and an energy normalization, the shape of 
the spectrum depends on $S$ and $\alpha$ only through a single parameter $\gamma$ (a 
property which is valid only at second order perturbation theory).

\section*{Appendix B: Linear spin-wave theory}

After applying the $\pi$ rotation $S_{i}^{x}\to  S_{i}^{x}$, $S_{i}^{y} \to- S^{y}_{i}$,
$S^{z}_{i} \to - S_{i}^{z}$ to the spins residing on the C sublattice, introducing the
Holstein-Primakoff representations
\begin{equation}
\begin{split}
S^{+}_{i} &  =  S^{x}_{i} + i S^{y}_{i} = \lt(\sqrt{2 S - n_{i}}\rt) a_{i}  \\
S^{-}_{i} & = S^{x}_{i} - i S^{y}_{i} =  a^{+}_{i} \sqrt{(2 S - n_{i})}~,\\
S^{z}_{i} & = S - n_{i}~,\qquad n_{i} = a^{+}_{i} a_{i}~,
\end{split}
\end{equation}
and truncating the Hamiltonian to terms quadratic in $a_{i}$, $a^{+}_{i}$ leads to
the spin-wave Hamiltonian:
\begin{equation}
\begin{split}
& {\cal H}_{\rm LSWT}  = - NS^{2} J_{zz} -\frac{1}{3} N S g_{z} \mu_{\rm B} H\\
& + S  J_{xy} \sum_{\langle i, j \rangle \in A, B} \lt(a^{+}_{i} a_{j} +
a^{+}_{j} a_{i} \rt) \\
& + S J_{xy} \sum_{i \in C} \sum_{j \in \text{n.n.}(i)} \lt(a_{i} a_{j} + a^{+}_{i} 
a^{+}_{j}\rt) \\
&  + g_{z} \mu_{\rm B}H \sum_{i \in A, B} n_{i} + (6 S J_{zz} - g_{z} \mu_{\rm B} H) 
\sum_{i \in C} n_{i}~.
\end{split}
\end{equation}

Here the sum $\sum_{\langle i, j \rangle \in A, B}$ runs over all pairs of bonds 
belonging to the honeycomb sublattice composed of the A and B sites.
The sum $\sum_{j \in \text{n.n.}(i)}$ runs over all nearest neighbours to the site $i$.

In momentum space, the Hamiltonian reads:
\begin{equation}
\begin{split}
{\cal H}_{\rm LSWT} & =  - NS(S + 1) J_{zz} -\frac{1}{6} N (2 S + 1) g_{z} \mu_{\rm B} H 
 \\
& + \frac{1}{2} \sum_{\bk} \begin{pmatrix}
       A^{+}_{\bk} A_{-\bk}
      \end{pmatrix}
\begin{vmatrix}
E_{\bk} & F_{\bk}\\
F_{\bk} & E_{\bk}
\end{vmatrix}
\begin{pmatrix}
A_{\bk} \\
A^{+}_{-\bk}
\end{pmatrix}~,
\end{split}
\end{equation}
where $A_{\bk} = (a_{{\rm A} \bk}, a_{{\rm B} \bk}, a_{{\rm C} \bk})$ and $A^{+}_{\bk}=
(a^{+}_{{\rm A} \bk}, a^{+}_{{\rm B}\bk} , a^{+}_{{\rm C}\bk})$ are the creation/annihilation
operators of plane wave states on the three sublattices and, $E_{\bk}$, $F_{\bk}$ are the $3
\times 3$ matrices
\begin{equation}
\begin{gathered}
E_{\bk} = \begin{vmatrix}
        g_{z} \mu_{\rm B} H & S J_{xy}f_{\bk} & 0 \\
        S J_{xy} f^{*}_{\bk} & g_{z} \mu_{\rm B} H & 0 \\
        0 & 0 & 6 S J_{zz} - g_{z} \mu_{\rm B} H
          \end{vmatrix}~, \\
F_{\bk}  =  S J_{xy} \begin{vmatrix}
           0 & 0 & f^{*}_{\bk}\\
           0 & 0 &  f_{\bk}\\
            f_{\bk} &  f^{*}_{\bk} & 0 \\
          \end{vmatrix}~.
\end{gathered}
\end{equation}

The spectrum can be found from the modulus of the eigenvalues of
\begin{equation} \label{eigenvalues_LSWT}
\begin{split} 
M(\bk) & = \Sigma^{z}  \cdot \begin{vmatrix}
E_{\bk} & F_{\bk}\\
F_{\bk} & E_{\bk}
\end{vmatrix}\\
& = \begin{vmatrix}
I & 0 \\
0 & -I 
\end{vmatrix}
\begin{vmatrix}
E_{\bk} & F_{\bk}\\
F_{\bk} & E_{\bk}
\end{vmatrix} = 
\begin{vmatrix}
E_{\bk} & F_{\bk}\\
-F_{\bk} & -E_{\bk}
\end{vmatrix}~,
\end{split}
\end{equation}
where $I$ is the $3 \times 3$ identity matrix.
Due to the structure of the matrices (which is a consequence of the $U(1)$ symmetry of the
plateau phase), $M(\bk)$ breaks into two decoupled $3 \times 3$ blocks, with opposite eigenvalues:
a block corresponding to the columns $1$, $2$, and $6$ and a block generated by the columns $3$,
$4$, and $5$.

As a result, the spectrum is given by the modulus of the eigenvalues of
\begin{equation}
\mathcal{M}^{\rm LSWT}(\bk) = g_{z} \mu_{\rm B} H I + S J_{zz} \begin{vmatrix}
                                                     0 & \alpha f_{\bk} & \alpha 
f^{*}_{\bk} \\
\alpha f^{*}_{\bk} & 0 & \alpha f_{\bk} \\
- \alpha f_{\bk} & - \alpha f^{*}_{\bk} & -6
\end{vmatrix}~,
\end{equation}
as in Eq.~\eqref{eigenvalues_LSWT} of the main text.

We note in passing that the linear spin-wave spectrum has a particularly simple 
analytical form along the line $k_{x} = 0$, on which $f_{\bk}$ is real-valued (this 
line corresponds to the $\Gamma$-M line in INS spectra).
The dispersions in LSWT for any $\alpha$ on this line are
\begin{equation}\label{LSWT_Gamma_M}
\begin{split}
\epsilon_{u-}(0, k_{y}) & = g_{z} \mu_{\rm B} H - S J_{xy} f_{\bk}~, \\
\epsilon_{u+}(0, k_{y}) & = g_{z} \mu_{\rm B} H - 3 S J_{zz} + \frac{S}{2} J_{xy}
f_{\bk} \\
& + \frac{S}{2} \sqrt{\lt(6 J_{zz} + J_{xy} f_{\bk}\rt)^{2} - 8 J_{xy}^{2} f^{2}_{\bk}}~,\\
\epsilon_{d}(0, k_{y}) & = 3 S J_{zz} - g_{z} \mu_{\rm B} H - \frac{S}{2} J_{xy} f_{\bk} \\
& + \frac{S}{2} \sqrt{\lt(6 J_{zz} + J_{xy} f_{\bk}\rt)^{2} - 8 J_{xy}^{2} f^{2}_{\bk}}~.
\end{split}
\end{equation}
with $f_{\bk} = 1 + 2 \cos(\sqrt{3} k_{y}/2)$.

In Ref.~\cite{Note1} it was shown that the experimental dispersions $\epsilon_{u\pm}(\bk)$ along the
$\Gamma$-K-M-$\Gamma$ lines in KCSO are well reproduced by the LSWT dispersions computed with
$\tilde{J}_{zz} = 0.68(6)$ meV, $\tilde{J}_{xy} = 0.25(1)$ meV, $\tilde{g}_{z}=8.34(6)$, although
the value of the renormalized parameter $\tilde{J}_{zz}$ is in strong disagreement with the
actual physical value of the exchange constant $J_{zz}$.
Although the fitting parameters are not consistent with the physical magnitude of the interactions,
we can use the results of the fitting of Ref.~\cite{Note1} to deduce the experimental ratio between
the widths $W_{+}$, $W_{-}$ of the two branches, which we calculate to be $W_{+}/W_{-} = 0.67(4)$.

\footnotetext[1]{See the Supplemental Materials of Ref.~\cite{zhu_prl_2024}.}

\footnotetext[2]{A. Mauri and F. Mila, 'Spin-wave expansion and magnon dispersion in the 1/3
plateau of the triangular XXZ antiferromagnet'. Zenodo, 2025.
doi: \href{https://doi.org/10.5281/zenodo.15270189}{https://doi.org/10.5281/zenodo.15270189}}


\begin{thebibliography}{26}%
\makeatletter
\providecommand \@ifxundefined [1]{%
 \@ifx{#1\undefined}
}%
\providecommand \@ifnum [1]{%
 \ifnum #1\expandafter \@firstoftwo
 \else \expandafter \@secondoftwo
 \fi
}%
\providecommand \@ifx [1]{%
 \ifx #1\expandafter \@firstoftwo
 \else \expandafter \@secondoftwo
 \fi
}%
\providecommand \natexlab [1]{#1}%
\providecommand \enquote  [1]{``#1''}%
\providecommand \bibnamefont  [1]{#1}%
\providecommand \bibfnamefont [1]{#1}%
\providecommand \citenamefont [1]{#1}%
\providecommand \href@noop [0]{\@secondoftwo}%
\providecommand \href [0]{\begingroup \@sanitize@url \@href}%
\providecommand \@href[1]{\@@startlink{#1}\@@href}%
\providecommand \@@href[1]{\endgroup#1\@@endlink}%
\providecommand \@sanitize@url [0]{\catcode `\\12\catcode `\$12\catcode
  `\&12\catcode `\#12\catcode `\^12\catcode `\_12\catcode `\%12\relax}%
\providecommand \@@startlink[1]{}%
\providecommand \@@endlink[0]{}%
\providecommand \url  [0]{\begingroup\@sanitize@url \@url }%
\providecommand \@url [1]{\endgroup\@href {#1}{\urlprefix }}%
\providecommand \urlprefix  [0]{URL }%
\providecommand \Eprint [0]{\href }%
\providecommand \doibase [0]{https://doi.org/}%
\providecommand \selectlanguage [0]{\@gobble}%
\providecommand \bibinfo  [0]{\@secondoftwo}%
\providecommand \bibfield  [0]{\@secondoftwo}%
\providecommand \translation [1]{[#1]}%
\providecommand \BibitemOpen [0]{}%
\providecommand \bibitemStop [0]{}%
\providecommand \bibitemNoStop [0]{.\EOS\space}%
\providecommand \EOS [0]{\spacefactor3000\relax}%
\providecommand \BibitemShut  [1]{\csname bibitem#1\endcsname}%
\let\auto@bib@innerbib\@empty
%</preamble>
\bibitem [{\citenamefont {Fazekas}\ and\ \citenamefont
  {Anderson}(1974)}]{fazekas_pm_1974}%
  \BibitemOpen
  \bibfield  {author} {\bibinfo {author} {\bibfnamefont {P.}~\bibnamefont
  {Fazekas}}\ and\ \bibinfo {author} {\bibfnamefont {P.~W.}\ \bibnamefont
  {Anderson}},\ }\bibfield  {title} {\bibinfo {title} {On the ground state
  properties of the anisotropic triangular antiferromagnet},\ }\href
  {https://doi.org/10.1080/14786439808206568} {\bibfield  {journal} {\bibinfo
  {journal} {Philos. Mag.}\ }\textbf {\bibinfo {volume} {30}},\ \bibinfo
  {pages} {423} (\bibinfo {year} {1974})}\BibitemShut {NoStop}%
\bibitem [{\citenamefont {Kleine}\ \emph
  {et~al.}(1992{\natexlab{a}})\citenamefont {Kleine}, \citenamefont
  {M\"{u}ller-Hartmann}, \citenamefont {Frahm},\ and\ \citenamefont
  {Fazekas}}]{kleine_zpb_1992}%
  \BibitemOpen
  \bibfield  {author} {\bibinfo {author} {\bibfnamefont {B.}~\bibnamefont
  {Kleine}}, \bibinfo {author} {\bibfnamefont {E.}~\bibnamefont
  {M\"{u}ller-Hartmann}}, \bibinfo {author} {\bibfnamefont {K.}~\bibnamefont
  {Frahm}},\ and\ \bibinfo {author} {\bibfnamefont {P.}~\bibnamefont
  {Fazekas}},\ }\bibfield  {title} {\bibinfo {title} {Spin-wave analysis of
  easy-axis quantum antiferromagnets on the triangular lattice},\ }\href
  {https://doi.org/10.1007/BF01308264} {\bibfield  {journal} {\bibinfo
  {journal} {Z. Phys. B Condens. Matter}\ }\textbf {\bibinfo {volume} {87}},\
  \bibinfo {pages} {103} (\bibinfo {year} {1992}{\natexlab{a}})}\BibitemShut
  {NoStop}%
\bibitem [{\citenamefont {Kleine}\ \emph
  {et~al.}(1992{\natexlab{b}})\citenamefont {Kleine}, \citenamefont {Fazekas},\
  and\ \citenamefont {M\"{u}ller-Hartmann}}]{kleine_zpb_1992b}%
  \BibitemOpen
  \bibfield  {author} {\bibinfo {author} {\bibfnamefont {B.}~\bibnamefont
  {Kleine}}, \bibinfo {author} {\bibfnamefont {P.}~\bibnamefont {Fazekas}},\
  and\ \bibinfo {author} {\bibfnamefont {E.}~\bibnamefont
  {M\"{u}ller-Hartmann}},\ }\bibfield  {title} {\bibinfo {title} {Perturbation
  theory for the triangular {Heisenberg} antiferromagnet with {Ising}-like
  anisotropy},\ }\href {https://doi.org/10.1007/BF01323734} {\bibfield
  {journal} {\bibinfo  {journal} {Z. Phys. B}\ }\textbf {\bibinfo {volume}
  {86}},\ \bibinfo {pages} {405} (\bibinfo {year}
  {1992}{\natexlab{b}})}\BibitemShut {NoStop}%
\bibitem [{\citenamefont {Sen}\ \emph {et~al.}(2008)\citenamefont {Sen},
  \citenamefont {Dutt}, \citenamefont {Damle},\ and\ \citenamefont
  {Moessner}}]{sen_prl_2008}%
  \BibitemOpen
  \bibfield  {author} {\bibinfo {author} {\bibfnamefont {A.}~\bibnamefont
  {Sen}}, \bibinfo {author} {\bibfnamefont {P.}~\bibnamefont {Dutt}}, \bibinfo
  {author} {\bibfnamefont {K.}~\bibnamefont {Damle}},\ and\ \bibinfo {author}
  {\bibfnamefont {R.}~\bibnamefont {Moessner}},\ }\bibfield  {title} {\bibinfo
  {title} {Variational wave-function study of the triangular lattice
  supersolid},\ }\href {https://doi.org/10.1103/PhysRevLett.100.147204}
  {\bibfield  {journal} {\bibinfo  {journal} {Phys. Rev. Lett.}\ }\textbf
  {\bibinfo {volume} {100}},\ \bibinfo {pages} {147204} (\bibinfo {year}
  {2008})}\BibitemShut {NoStop}%
\bibitem [{\citenamefont {Wang}\ \emph {et~al.}(2009)\citenamefont {Wang},
  \citenamefont {Pollmann},\ and\ \citenamefont {Vishwanath}}]{wang_prl_2009}%
  \BibitemOpen
  \bibfield  {author} {\bibinfo {author} {\bibfnamefont {F.}~\bibnamefont
  {Wang}}, \bibinfo {author} {\bibfnamefont {F.}~\bibnamefont {Pollmann}},\
  and\ \bibinfo {author} {\bibfnamefont {A.}~\bibnamefont {Vishwanath}},\
  }\bibfield  {title} {\bibinfo {title} {Extended supersolid phase of
  frustrated hard-core bosons on a triangular lattice},\ }\href
  {https://doi.org/10.1103/PhysRevLett.102.017203} {\bibfield  {journal}
  {\bibinfo  {journal} {Phys. Rev. Lett.}\ }\textbf {\bibinfo {volume} {102}},\
  \bibinfo {pages} {017203} (\bibinfo {year} {2009})}\BibitemShut {NoStop}%
\bibitem [{\citenamefont {Jiang}\ \emph {et~al.}(2009)\citenamefont {Jiang},
  \citenamefont {Weng}, \citenamefont {Weng}, \citenamefont {Sheng},\ and\
  \citenamefont {Balents}}]{jiang_prb_2009}%
  \BibitemOpen
  \bibfield  {author} {\bibinfo {author} {\bibfnamefont {H.~C.}\ \bibnamefont
  {Jiang}}, \bibinfo {author} {\bibfnamefont {M.~Q.}\ \bibnamefont {Weng}},
  \bibinfo {author} {\bibfnamefont {Z.~Y.}\ \bibnamefont {Weng}}, \bibinfo
  {author} {\bibfnamefont {D.~N.}\ \bibnamefont {Sheng}},\ and\ \bibinfo
  {author} {\bibfnamefont {L.}~\bibnamefont {Balents}},\ }\bibfield  {title}
  {\bibinfo {title} {Supersolid order of frustrated hard-core bosons in a
  triangular lattice system},\ }\href
  {https://doi.org/10.1103/PhysRevB.79.020409} {\bibfield  {journal} {\bibinfo
  {journal} {Phys. Rev. B}\ }\textbf {\bibinfo {volume} {79}},\ \bibinfo
  {pages} {020409(R)} (\bibinfo {year} {2009})}\BibitemShut {NoStop}%
\bibitem [{\citenamefont {Heidarian}\ and\ \citenamefont
  {Paramekanti}(2010)}]{heidarian_prl_2010}%
  \BibitemOpen
  \bibfield  {author} {\bibinfo {author} {\bibfnamefont {D.}~\bibnamefont
  {Heidarian}}\ and\ \bibinfo {author} {\bibfnamefont {A.}~\bibnamefont
  {Paramekanti}},\ }\bibfield  {title} {\bibinfo {title} {Supersolidity in the
  triangular lattice spin-$1/2$ {XXZ} model: a variational perspective},\
  }\href {https://doi.org/10.1103/PhysRevLett.104.015301} {\bibfield  {journal}
  {\bibinfo  {journal} {Phys. Rev. Lett.}\ }\textbf {\bibinfo {volume} {104}},\
  \bibinfo {pages} {015301} (\bibinfo {year} {2010})}\BibitemShut {NoStop}%
\bibitem [{\citenamefont {Zhang}\ \emph {et~al.}(2011)\citenamefont {Zhang},
  \citenamefont {Dillenschneider}, \citenamefont {Yu},\ and\ \citenamefont
  {Eggert}}]{zhang_prb_2011}%
  \BibitemOpen
  \bibfield  {author} {\bibinfo {author} {\bibfnamefont {X.-F.}\ \bibnamefont
  {Zhang}}, \bibinfo {author} {\bibfnamefont {R.}~\bibnamefont
  {Dillenschneider}}, \bibinfo {author} {\bibfnamefont {Y.}~\bibnamefont
  {Yu}},\ and\ \bibinfo {author} {\bibfnamefont {S.}~\bibnamefont {Eggert}},\
  }\bibfield  {title} {\bibinfo {title} {Supersolid phase transitions for
  hard-core bosons on a triangular lattice},\ }\href
  {https://doi.org/10.1103/PhysRevB.84.174515} {\bibfield  {journal} {\bibinfo
  {journal} {Phys. Rev. B}\ }\textbf {\bibinfo {volume} {84}},\ \bibinfo
  {pages} {174515} (\bibinfo {year} {2011})}\BibitemShut {NoStop}%
\bibitem [{\citenamefont {Yamamoto}\ \emph {et~al.}(2014)\citenamefont
  {Yamamoto}, \citenamefont {Marmorini},\ and\ \citenamefont
  {Danshita}}]{yamamoto_prl_2014}%
  \BibitemOpen
  \bibfield  {author} {\bibinfo {author} {\bibfnamefont {D.}~\bibnamefont
  {Yamamoto}}, \bibinfo {author} {\bibfnamefont {G.}~\bibnamefont
  {Marmorini}},\ and\ \bibinfo {author} {\bibfnamefont {I.}~\bibnamefont
  {Danshita}},\ }\bibfield  {title} {\bibinfo {title} {Quantum phase diagram of
  the triangular-lattice {XXZ} model in a magnetic field},\ }\href
  {https://doi.org/10.1103/PhysRevLett.112.127203} {\bibfield  {journal}
  {\bibinfo  {journal} {Phys. Rev. Lett.}\ }\textbf {\bibinfo {volume} {112}},\
  \bibinfo {pages} {127203} (\bibinfo {year} {2014})}\BibitemShut {NoStop}%
\bibitem [{\citenamefont {Sellmann}\ \emph {et~al.}(2015)\citenamefont
  {Sellmann}, \citenamefont {Zhang},\ and\ \citenamefont
  {Eggert}}]{sellmann_prb_2015}%
  \BibitemOpen
  \bibfield  {author} {\bibinfo {author} {\bibfnamefont {D.}~\bibnamefont
  {Sellmann}}, \bibinfo {author} {\bibfnamefont {X.-F.}\ \bibnamefont
  {Zhang}},\ and\ \bibinfo {author} {\bibfnamefont {S.}~\bibnamefont
  {Eggert}},\ }\bibfield  {title} {\bibinfo {title} {Phase diagram of the
  antiferromagnetic {XXZ} model on the triangular lattice},\ }\href
  {https://doi.org/10.1103/PhysRevB.91.081104} {\bibfield  {journal} {\bibinfo
  {journal} {Phys. Rev. B}\ }\textbf {\bibinfo {volume} {91}},\ \bibinfo
  {pages} {081104(R)} (\bibinfo {year} {2015})}\BibitemShut {NoStop}%
\bibitem [{\citenamefont {Yamamoto}\ \emph {et~al.}(2019)\citenamefont
  {Yamamoto}, \citenamefont {Marmorini}, \citenamefont {Tabata}, \citenamefont
  {Sakakura},\ and\ \citenamefont {Danshita}}]{yamamoto_prb_2019}%
  \BibitemOpen
  \bibfield  {author} {\bibinfo {author} {\bibfnamefont {D.}~\bibnamefont
  {Yamamoto}}, \bibinfo {author} {\bibfnamefont {G.}~\bibnamefont {Marmorini}},
  \bibinfo {author} {\bibfnamefont {M.}~\bibnamefont {Tabata}}, \bibinfo
  {author} {\bibfnamefont {K.}~\bibnamefont {Sakakura}},\ and\ \bibinfo
  {author} {\bibfnamefont {I.}~\bibnamefont {Danshita}},\ }\bibfield  {title}
  {\bibinfo {title} {Magnetism driven by the interplay of fluctuations and
  frustration in the easy-axis triangular {XXZ} model with transverse fields},\
  }\href {https://doi.org/10.1103/PhysRevB.100.140410} {\bibfield  {journal}
  {\bibinfo  {journal} {Phys. Rev. B}\ }\textbf {\bibinfo {volume} {100}},\
  \bibinfo {pages} {140410} (\bibinfo {year} {2019})}\BibitemShut {NoStop}%
\bibitem [{\citenamefont {Xu}\ \emph {et~al.}(2024)\citenamefont {Xu},
  \citenamefont {Hasik}, \citenamefont {Ponsioen},\ and\ \citenamefont
  {Nevidomskyy}}]{xu_arxiv_2024}%
  \BibitemOpen
  \bibfield  {author} {\bibinfo {author} {\bibfnamefont {Y.}~\bibnamefont
  {Xu}}, \bibinfo {author} {\bibfnamefont {J.}~\bibnamefont {Hasik}}, \bibinfo
  {author} {\bibfnamefont {B.}~\bibnamefont {Ponsioen}},\ and\ \bibinfo
  {author} {\bibfnamefont {A.~H.}\ \bibnamefont {Nevidomskyy}},\ }\bibfield
  {title} {\bibinfo {title} {Simulating spin dynamics of supersolid states in a
  quantum {Ising} magnet},\ }\bibfield  {journal} {\bibinfo  {journal}
  {arXiv:2405.05151}\ }\href {https://doi.org/10.48550/arXiv.2405.05151}
  {10.48550/arXiv.2405.05151} (\bibinfo {year} {2024})\BibitemShut {NoStop}%
\bibitem [{\citenamefont {Ulaga}\ \emph
  {et~al.}(2024{\natexlab{a}})\citenamefont {Ulaga}, \citenamefont {Kokalj},\
  and\ \citenamefont {Prelov\v{s}ek}}]{ulaga_arxiv_2024}%
  \BibitemOpen
  \bibfield  {author} {\bibinfo {author} {\bibfnamefont {M.}~\bibnamefont
  {Ulaga}}, \bibinfo {author} {\bibfnamefont {J.}~\bibnamefont {Kokalj}},\ and\
  \bibinfo {author} {\bibfnamefont {P.}~\bibnamefont {Prelov\v{s}ek}},\
  }\bibfield  {title} {\bibinfo {title} {Easy-axis {Heisenberg} model on the
  triangular lattice: from supersolid to gapped solid},\ }\bibfield  {journal}
  {\bibinfo  {journal} {arXiv:2408.05034}\ }\href
  {https://doi.org/10.48550/arXiv.2408.05034} {10.48550/arXiv.2408.05034}
  (\bibinfo {year} {2024}{\natexlab{a}})\BibitemShut {NoStop}%
\bibitem [{\citenamefont {Ulaga}\ \emph
  {et~al.}(2024{\natexlab{b}})\citenamefont {Ulaga}, \citenamefont {Kokalj},
  \citenamefont {Wietek}, \citenamefont {Zorko},\ and\ \citenamefont
  {Prelov\v{s}ek}}]{ulaga_prb_2024}%
  \BibitemOpen
  \bibfield  {author} {\bibinfo {author} {\bibfnamefont {M.}~\bibnamefont
  {Ulaga}}, \bibinfo {author} {\bibfnamefont {J.}~\bibnamefont {Kokalj}},
  \bibinfo {author} {\bibfnamefont {A.}~\bibnamefont {Wietek}}, \bibinfo
  {author} {\bibfnamefont {A.}~\bibnamefont {Zorko}},\ and\ \bibinfo {author}
  {\bibfnamefont {P.}~\bibnamefont {Prelov\v{s}ek}},\ }\bibfield  {title}
  {\bibinfo {title} {Finite-temperature properties of the easy-axis
  {Heisenberg} model on frustrated lattices},\ }\href
  {https://doi.org/10.1103/PhysRevB.109.035110} {\bibfield  {journal} {\bibinfo
   {journal} {Phys. Rev. B}\ }\textbf {\bibinfo {volume} {109}},\ \bibinfo
  {pages} {035110} (\bibinfo {year} {2024}{\natexlab{b}})}\BibitemShut
  {NoStop}%
\bibitem [{\citenamefont {Gao}\ \emph {et~al.}(2022)\citenamefont {Gao},
  \citenamefont {Fan}, \citenamefont {Li}, \citenamefont {Yang}, \citenamefont
  {Zeng}, \citenamefont {Sheng}, \citenamefont {Zhong}, \citenamefont {Qi},
  \citenamefont {Wan},\ and\ \citenamefont {Li}}]{gao_npj_2022}%
  \BibitemOpen
  \bibfield  {author} {\bibinfo {author} {\bibfnamefont {Y.}~\bibnamefont
  {Gao}}, \bibinfo {author} {\bibfnamefont {Y.-C.}\ \bibnamefont {Fan}},
  \bibinfo {author} {\bibfnamefont {H.}~\bibnamefont {Li}}, \bibinfo {author}
  {\bibfnamefont {F.}~\bibnamefont {Yang}}, \bibinfo {author} {\bibfnamefont
  {X.-T.}\ \bibnamefont {Zeng}}, \bibinfo {author} {\bibfnamefont {X.-L.}\
  \bibnamefont {Sheng}}, \bibinfo {author} {\bibfnamefont {R.}~\bibnamefont
  {Zhong}}, \bibinfo {author} {\bibfnamefont {Y.}~\bibnamefont {Qi}}, \bibinfo
  {author} {\bibfnamefont {Y.}~\bibnamefont {Wan}},\ and\ \bibinfo {author}
  {\bibfnamefont {W.}~\bibnamefont {Li}},\ }\bibfield  {title} {\bibinfo
  {title} {Spin supersolidity in nearly ideal easy-axis triangular quantum
  antiferromagnet {Na$_{2}$BaCo(PO$_{4}$)$_{2}$}},\ }\href
  {https://doi.org/10.1038/s41535-022-00500-3} {\bibfield  {journal} {\bibinfo
  {journal} {npj Quantum Mater.}\ }\textbf {\bibinfo {volume} {7}},\ \bibinfo
  {pages} {89} (\bibinfo {year} {2022})}\BibitemShut {NoStop}%
\bibitem [{\citenamefont {Zhong}\ \emph {et~al.}(2019)\citenamefont {Zhong},
  \citenamefont {Guo}, \citenamefont {Xu}, \citenamefont {Xu},\ and\
  \citenamefont {Cava}}]{zhong_pnas_2019}%
  \BibitemOpen
  \bibfield  {author} {\bibinfo {author} {\bibfnamefont {R.}~\bibnamefont
  {Zhong}}, \bibinfo {author} {\bibfnamefont {S.}~\bibnamefont {Guo}}, \bibinfo
  {author} {\bibfnamefont {G.}~\bibnamefont {Xu}}, \bibinfo {author}
  {\bibfnamefont {Z.}~\bibnamefont {Xu}},\ and\ \bibinfo {author}
  {\bibfnamefont {R.~J.}\ \bibnamefont {Cava}},\ }\bibfield  {title} {\bibinfo
  {title} {Strong quantum fluctuations in a quantum spin liquid candidate with
  a {Co}-based triangular lattice},\ }\href
  {https://doi.org/10.1073/pnas.1906483116} {\bibfield  {journal} {\bibinfo
  {journal} {Proc. Natl. Acad. Sci. U.S.A.}\ }\textbf {\bibinfo {volume}
  {116}},\ \bibinfo {pages} {14505} (\bibinfo {year} {2019})}\BibitemShut
  {NoStop}%
\bibitem [{\citenamefont {Jia}\ \emph {et~al.}(2024)\citenamefont {Jia},
  \citenamefont {Ma}, \citenamefont {Wang},\ and\ \citenamefont
  {Chen}}]{jia_prr_2024}%
  \BibitemOpen
  \bibfield  {author} {\bibinfo {author} {\bibfnamefont {H.}~\bibnamefont
  {Jia}}, \bibinfo {author} {\bibfnamefont {B.}~\bibnamefont {Ma}}, \bibinfo
  {author} {\bibfnamefont {Z.}~\bibnamefont {Wang}},\ and\ \bibinfo {author}
  {\bibfnamefont {G.}~\bibnamefont {Chen}},\ }\bibfield  {title} {\bibinfo
  {title} {Quantum spin supersolid as a precursory {Dirac} spin liquid in a
  triangular lattice antiferromagnet},\ }\href
  {https://doi.org/10.1103/PhysRevResearch.6.033031} {\bibfield  {journal}
  {\bibinfo  {journal} {Phys. Rev. Research}\ }\textbf {\bibinfo {volume}
  {6}},\ \bibinfo {pages} {033031} (\bibinfo {year} {2024})}\BibitemShut
  {NoStop}%
\bibitem [{\citenamefont {Chi}\ \emph {et~al.}(2024)\citenamefont {Chi},
  \citenamefont {Hu}, \citenamefont {Liao},\ and\ \citenamefont
  {Xiang}}]{chi_prb_2024}%
  \BibitemOpen
  \bibfield  {author} {\bibinfo {author} {\bibfnamefont {R.}~\bibnamefont
  {Chi}}, \bibinfo {author} {\bibfnamefont {J.}~\bibnamefont {Hu}}, \bibinfo
  {author} {\bibfnamefont {H.-J.}\ \bibnamefont {Liao}},\ and\ \bibinfo
  {author} {\bibfnamefont {T.}~\bibnamefont {Xiang}},\ }\bibfield  {title}
  {\bibinfo {title} {Dynamical spectra of spin supersolid states in triangular
  antiferromagnets},\ }\href {https://doi.org/10.48550/arXiv.2404.14163}
  {\bibfield  {journal} {\bibinfo  {journal} {Phys. Rev. B}\ }\textbf {\bibinfo
  {volume} {110}},\ \bibinfo {pages} {L180404} (\bibinfo {year}
  {2024})}\BibitemShut {NoStop}%
\bibitem [{\citenamefont {Sheng}\ \emph {et~al.}(2024)\citenamefont {Sheng},
  \citenamefont {Wang}, \citenamefont {Jiang}, \citenamefont {Ge},
  \citenamefont {Zhao}, \citenamefont {Li}, \citenamefont {Kofu}, \citenamefont
  {Yu}, \citenamefont {Zhu}, \citenamefont {Mei}, \citenamefont {Wang},\ and\
  \citenamefont {Wu}}]{sheng_arxiv_2024}%
  \BibitemOpen
  \bibfield  {author} {\bibinfo {author} {\bibfnamefont {J.}~\bibnamefont
  {Sheng}}, \bibinfo {author} {\bibfnamefont {L.}~\bibnamefont {Wang}},
  \bibinfo {author} {\bibfnamefont {W.}~\bibnamefont {Jiang}}, \bibinfo
  {author} {\bibfnamefont {H.}~\bibnamefont {Ge}}, \bibinfo {author}
  {\bibfnamefont {N.}~\bibnamefont {Zhao}}, \bibinfo {author} {\bibfnamefont
  {T.}~\bibnamefont {Li}}, \bibinfo {author} {\bibfnamefont {M.}~\bibnamefont
  {Kofu}}, \bibinfo {author} {\bibfnamefont {D.}~\bibnamefont {Yu}}, \bibinfo
  {author} {\bibfnamefont {W.}~\bibnamefont {Zhu}}, \bibinfo {author}
  {\bibfnamefont {J.-W.}\ \bibnamefont {Mei}}, \bibinfo {author} {\bibfnamefont
  {Z.}~\bibnamefont {Wang}},\ and\ \bibinfo {author} {\bibfnamefont
  {L.}~\bibnamefont {Wu}},\ }\bibfield  {title} {\bibinfo {title} {Continuum of
  spin excitations in an ordered magnet},\ }\bibfield  {journal} {\bibinfo
  {journal} {arXiv:2402.077730}\ }\href
  {https://doi.org/10.48550/arXiv.2402.07730} {10.48550/arXiv.2402.07730}
  (\bibinfo {year} {2024})\BibitemShut {NoStop}%
\bibitem [{\citenamefont {Arh}\ \emph {et~al.}(2022)\citenamefont {Arh},
  \citenamefont {Sana}, \citenamefont {Pregelj}, \citenamefont {Khuntia},
  \citenamefont {Jagli\v{c}i\'{c}}, \citenamefont {Le}, \citenamefont {Biswas},
  \citenamefont {Manuel}, \citenamefont {Mangin-Thro}, \citenamefont
  {Ozarowski},\ and\ \citenamefont {Zorko}}]{arh_nm_2022}%
  \BibitemOpen
  \bibfield  {author} {\bibinfo {author} {\bibfnamefont {T.}~\bibnamefont
  {Arh}}, \bibinfo {author} {\bibfnamefont {B.}~\bibnamefont {Sana}}, \bibinfo
  {author} {\bibfnamefont {M.}~\bibnamefont {Pregelj}}, \bibinfo {author}
  {\bibfnamefont {P.}~\bibnamefont {Khuntia}}, \bibinfo {author} {\bibfnamefont
  {Z.}~\bibnamefont {Jagli\v{c}i\'{c}}}, \bibinfo {author} {\bibfnamefont
  {M.~D.}\ \bibnamefont {Le}}, \bibinfo {author} {\bibfnamefont {P.~K.}\
  \bibnamefont {Biswas}}, \bibinfo {author} {\bibfnamefont {P.}~\bibnamefont
  {Manuel}}, \bibinfo {author} {\bibfnamefont {L.}~\bibnamefont {Mangin-Thro}},
  \bibinfo {author} {\bibfnamefont {A.}~\bibnamefont {Ozarowski}},\ and\
  \bibinfo {author} {\bibfnamefont {A.}~\bibnamefont {Zorko}},\ }\bibfield
  {title} {\bibinfo {title} {The {Ising} triangular-lattice antiferromagnet
  neodymium heptatantalate as a quantum spin liquid candidate},\ }\href
  {https://doi.org/10.1038/s41563-021-01169-y} {\bibfield  {journal} {\bibinfo
  {journal} {Nat. Mater.}\ }\textbf {\bibinfo {volume} {21}},\ \bibinfo {pages}
  {416} (\bibinfo {year} {2022})}\BibitemShut {NoStop}%
\bibitem [{\citenamefont {Zhong}\ \emph {et~al.}(2020)\citenamefont {Zhong},
  \citenamefont {Guo},\ and\ \citenamefont {Cava}}]{zhong_prm_2020}%
  \BibitemOpen
  \bibfield  {author} {\bibinfo {author} {\bibfnamefont {R.}~\bibnamefont
  {Zhong}}, \bibinfo {author} {\bibfnamefont {S.}~\bibnamefont {Guo}},\ and\
  \bibinfo {author} {\bibfnamefont {R.~J.}\ \bibnamefont {Cava}},\ }\bibfield
  {title} {\bibinfo {title} {Frustrated magnetism in the layered triangular
  lattice materials {K$_{2}$Co(SeO$_{3}$)$_{2}$} and
  {Rb$_{2}$Co(SeO$_{3}$)$_{2}$}},\ }\href
  {https://doi.org/10.1103/PhysRevMaterials.4.084406} {\bibfield  {journal}
  {\bibinfo  {journal} {Phys. Rev. Mater.}\ }\textbf {\bibinfo {volume} {4}},\
  \bibinfo {pages} {084406} (\bibinfo {year} {2020})}\BibitemShut {NoStop}%
\bibitem [{\citenamefont {Zhu}\ \emph {et~al.}(2024)\citenamefont {Zhu},
  \citenamefont {Romerio}, \citenamefont {Steiger}, \citenamefont {Nabi},
  \citenamefont {Murai}, \citenamefont {Ohira-Kawamura}, \citenamefont
  {Povarov}, \citenamefont {Skourski}, \citenamefont {Sibille}, \citenamefont
  {Keller}, \citenamefont {Yan}, \citenamefont {Gvasaliya},\ and\ \citenamefont
  {Zheludev}}]{zhu_prl_2024}%
  \BibitemOpen
  \bibfield  {author} {\bibinfo {author} {\bibfnamefont {M.}~\bibnamefont
  {Zhu}}, \bibinfo {author} {\bibfnamefont {V.}~\bibnamefont {Romerio}},
  \bibinfo {author} {\bibfnamefont {N.}~\bibnamefont {Steiger}}, \bibinfo
  {author} {\bibfnamefont {S.~D.}\ \bibnamefont {Nabi}}, \bibinfo {author}
  {\bibfnamefont {N.}~\bibnamefont {Murai}}, \bibinfo {author} {\bibfnamefont
  {S.}~\bibnamefont {Ohira-Kawamura}}, \bibinfo {author} {\bibfnamefont
  {K.~Y.}\ \bibnamefont {Povarov}}, \bibinfo {author} {\bibfnamefont
  {Y.}~\bibnamefont {Skourski}}, \bibinfo {author} {\bibfnamefont
  {R.}~\bibnamefont {Sibille}}, \bibinfo {author} {\bibfnamefont
  {L.}~\bibnamefont {Keller}}, \bibinfo {author} {\bibfnamefont
  {Z.}~\bibnamefont {Yan}}, \bibinfo {author} {\bibfnamefont {S.}~\bibnamefont
  {Gvasaliya}},\ and\ \bibinfo {author} {\bibfnamefont {A.}~\bibnamefont
  {Zheludev}},\ }\bibfield  {title} {\bibinfo {title} {Continuum excitations in
  a spin supersolid on a triangular lattice},\ }\href
  {https://doi.org/10.1103/PhysRevLett.133.186704} {\bibfield  {journal}
  {\bibinfo  {journal} {Phys. Rev. Lett.}\ }\textbf {\bibinfo {volume} {133}},\
  \bibinfo {pages} {186704} (\bibinfo {year} {2024})}\BibitemShut {NoStop}%
\bibitem [{\citenamefont {Chen}\ \emph {et~al.}(2024)\citenamefont {Chen},
  \citenamefont {Ghasemi}, \citenamefont {Zhang}, \citenamefont {Shi},
  \citenamefont {Tagay}, \citenamefont {Chen}, \citenamefont {Choi},
  \citenamefont {Jaime}, \citenamefont {Lee}, \citenamefont {Hao},
  \citenamefont {Cao}, \citenamefont {Winn}, \citenamefont {Zhong},
  \citenamefont {Xu}, \citenamefont {Armitage}, \citenamefont {Cava},\ and\
  \citenamefont {Broholm}}]{chen_arxiv_2024}%
  \BibitemOpen
  \bibfield  {author} {\bibinfo {author} {\bibfnamefont {T.}~\bibnamefont
  {Chen}}, \bibinfo {author} {\bibfnamefont {A.}~\bibnamefont {Ghasemi}},
  \bibinfo {author} {\bibfnamefont {J.}~\bibnamefont {Zhang}}, \bibinfo
  {author} {\bibfnamefont {L.}~\bibnamefont {Shi}}, \bibinfo {author}
  {\bibfnamefont {Z.}~\bibnamefont {Tagay}}, \bibinfo {author} {\bibfnamefont
  {L.}~\bibnamefont {Chen}}, \bibinfo {author} {\bibfnamefont {E.-S.}\
  \bibnamefont {Choi}}, \bibinfo {author} {\bibfnamefont {M.}~\bibnamefont
  {Jaime}}, \bibinfo {author} {\bibfnamefont {M.}~\bibnamefont {Lee}}, \bibinfo
  {author} {\bibfnamefont {Y.}~\bibnamefont {Hao}}, \bibinfo {author}
  {\bibfnamefont {H.}~\bibnamefont {Cao}}, \bibinfo {author} {\bibfnamefont
  {B.}~\bibnamefont {Winn}}, \bibinfo {author} {\bibfnamefont {R.}~\bibnamefont
  {Zhong}}, \bibinfo {author} {\bibfnamefont {X.}~\bibnamefont {Xu}}, \bibinfo
  {author} {\bibfnamefont {N.~P.}\ \bibnamefont {Armitage}}, \bibinfo {author}
  {\bibfnamefont {R.}~\bibnamefont {Cava}},\ and\ \bibinfo {author}
  {\bibfnamefont {C.}~\bibnamefont {Broholm}},\ }\bibfield  {title} {\bibinfo
  {title} {Phase diagram and spectroscopic evidence of supersolids in quantum
  {Ising} magnet {K$_{2}$Co(SeO$_{3}$)$_{2}$}},\ }\bibfield  {journal}
  {\bibinfo  {journal} {arXiv:2402.15869}\ }\href
  {https://doi.org/10.48550/arXiv.2402.15869} {10.48550/arXiv.2402.15869}
  (\bibinfo {year} {2024})\BibitemShut {NoStop}%
\bibitem [{\citenamefont {Kamiya}\ \emph {et~al.}(2018)\citenamefont {Kamiya},
  \citenamefont {Ge}, \citenamefont {Hong}, \citenamefont {Qiu}, \citenamefont
  {Quintero-Castro}, \citenamefont {Lu}, \citenamefont {Cao}, \citenamefont
  {Matsuda}, \citenamefont {Choi}, \citenamefont {Batista}, \citenamefont
  {Mourigal}, \citenamefont {Zhou},\ and\ \citenamefont {Ma}}]{kamiya_nc_2018}%
  \BibitemOpen
  \bibfield  {author} {\bibinfo {author} {\bibfnamefont {Y.}~\bibnamefont
  {Kamiya}}, \bibinfo {author} {\bibfnamefont {L.}~\bibnamefont {Ge}}, \bibinfo
  {author} {\bibfnamefont {T.}~\bibnamefont {Hong}}, \bibinfo {author}
  {\bibfnamefont {Y.}~\bibnamefont {Qiu}}, \bibinfo {author} {\bibfnamefont
  {D.~L.}\ \bibnamefont {Quintero-Castro}}, \bibinfo {author} {\bibfnamefont
  {Z.}~\bibnamefont {Lu}}, \bibinfo {author} {\bibfnamefont {H.~B.}\
  \bibnamefont {Cao}}, \bibinfo {author} {\bibfnamefont {M.}~\bibnamefont
  {Matsuda}}, \bibinfo {author} {\bibfnamefont {E.~S.}\ \bibnamefont {Choi}},
  \bibinfo {author} {\bibfnamefont {C.~D.}\ \bibnamefont {Batista}}, \bibinfo
  {author} {\bibfnamefont {M.}~\bibnamefont {Mourigal}}, \bibinfo {author}
  {\bibfnamefont {H.~D.}\ \bibnamefont {Zhou}},\ and\ \bibinfo {author}
  {\bibfnamefont {J.}~\bibnamefont {Ma}},\ }\bibfield  {title} {\bibinfo
  {title} {The nature of spin excitations in the one-third magnetization
  plateau phase of {Ba$_{3}$CoSb$_{2}$O$_{9}$}},\ }\href
  {https://doi.org/10.1038/s41467-018-04914-1} {\bibfield  {journal} {\bibinfo
  {journal} {Nature Commun.}\ }\textbf {\bibinfo {volume} {9}},\ \bibinfo
  {pages} {2666} (\bibinfo {year} {2018})}\BibitemShut {NoStop}%
\bibitem [{Note1()}]{Note1}%
  \BibitemOpen
  \bibinfo {note} {See the Supplemental Materials of Ref.~\cite
  {zhu_prl_2024}.}\BibitemShut {Stop}%
\bibitem [{Note2()}]{Note2}%
  \BibitemOpen
  \bibinfo {note} {A. Mauri and F. Mila, 'Spin-wave expansion and magnon
  dispersion in the 1/3 plateau of the triangular XXZ antiferromagnet'. Zenodo,
  2025. doi: \protect \href
  {https://doi.org/10.5281/zenodo.15270189}{https://doi.org/10.5281/zenodo.15270189}}\BibitemShut
  {NoStop}%
\end{thebibliography}
\end{document}